# Protecting others vs. protecting yourself against ballistic droplets: Quantification by stain patterns

V. Márquez-Alvarez[1], J. Amigó-Vera[1], A. Rivera[2], A. J. Batista-Leyva[3,1], E. Altshuler[1†]

[1] Group of Complex Systems and Statistical Physics, Physics Faculty, University of Havana, 10400 Havana, Cuba.
[2] Zeolite Engineering Laboratory, Institute of Materials and Reagents (IMRE), University of Havana, 10400 Havana, Cuba.
[3] Instituto Superior de Ciencia y Tecnología Aplicadas (InSTEC), University of Havana, 10400 Havana, Cuba.
[†] ealtshuler@fisica.uh.cu

**Abstract**

It is often accepted *a priori* that a face mask worn by an infected subject is effective to avoid the spreading of a respiratory disease, while a healthy person is not necessarily well protected when wearing the mask. Using a frugal stain technique, we quantify the "ballistic" droplets reaching a receptor from a jet-emitting source which mimics a coughing, sneezing or talking human –in real life, such droplets may host active SARS-CoV-2 virus able to replicate in the nasopharynx. We demonstrate that materials often used in home-made face masks block most of the droplets. We also show quantitatively that less liquid carried by ballistic droplets reaches a receptor when a blocking material is deployed near the source than when located near the receptor, which supports the paradigm that your face mask does protect you, but protects others even better than you.

## INTRODUCTION

More than one year has passed since start of the COVID-19 outbreak. From the early beginnings, face masks have had a key role in cutting off the transmission together with other preventive measures such as quarantines, physical distancing and hand hygiene. Many studies assessed masks efficacy [Chu2020, Agrawal2021, Verma2020, Lyu2020, Howard2021, Xi2020, Robinson2021, Maggiolo2021, Dbouk2020, Fischer2020, Klompas2020, Mittal2020, Mirikar2021, Sickbert-Bennett2020, Akhtar2020, Davies2013, Leung2020, Kumar2020, Shah2021], which compelled almost every health agency, including WHO [WHO2020], CDC [CDC2021] and ECDC [ECDC2021], to recommend the use of face masks in certain settings.

By the end of 2020, the first vaccines were available [Zimmer2021], which has concentrated a great deal of attention. However, by the time of writing this paper, less than 30% of the world population has been fully vaccinated [OWDVaccine2021], and the number goes below 1% in the case of low-income countries, representing a total population of 665 million people [WorldBank2021]. In fact, a recent article [Irwin2021] declares "unlikely" the fact that people from low-income countries will be fully vaccinated by the end of 2022.

Moreover, evidence suggests that the Delta variant of the SARS-CoV-2 is substantially more transmissible than previous strains of the virus [Campbell2021]. Also, vaccinated people infected with the Delta strain could carry the same amount of viral load than an unvaccinated person, meaning that they can also transmit the virus [Riemersma2021]. These findings have recently made CDC to update its guidance and recommend wearing a mask in public indoor places, in areas of substantial or high transmission risk, even if they are fully vaccinated [CDCVaccinated2021].

So, in the present context face masks still play a central role in the fight against the pandemic.

Evidence suggests that SARS-CoV-2 is transmitted mainly through direct exposure to respiratory droplets carrying the virus [Meyerowitz2021]. This respiratory transmission route is usually split in two ways. The first –that we will call *ballistic transmission*– occurs by the emission of large droplets of fluids as infected individuals sneeze, cough, sing, talk or breath. Those droplets can get in touch with the mucous membranes (eyes, nose or mouth) of a susceptible person and infect her. Otherwise they fall to the ground within 1-2 meters in the horizontal direction. The second –*aerosol transmission*– is linked to fine droplets, which, thanks to Brownian motion, can be suspended in the air for hours and easily travel with air currents. While the separation line between the two mechanisms is difficult to define [Meyerowitz2021, Bahl2020], in many situations it is possible to know what mechanism is dominating.

The relevance of aerosol transmission has been a subject of heated debate during the pandemic. Some studies conclude that aerosol transmission is plausible [Bahl2020, Morawskaa2020, Anderson2020, Asadi2020, Bourouiba2021], while others agree, but argue that it involves low risk [Somsen2020, Smith2020], since before the 2-meter ($\approx$6-feet) length scale, large droplets carry more viral load than airborne particles. Beyond that distance, aerosol droplets dilute in the air, and are easily carried by air currents (except in poorly ventilated places where increasing viral load concentration could make infection possible). In fact, the strong dependence of COVID-19 infection risk with people proximity suggests that *ballistic* transmission is more relevant than aerosol transmission [Meyerowitz2021].

The present study has two objectives (1) Testing experimentally the hypothesis that a face mask protects others better than the wearer (i.e., it works better at source control than at target protection) and (2) Quantifying the difference between exclusive source mask-wearing and exclusive target mask-wearing.

Addressing these questions has gained a renewed relevance, as some countries have lifted the mandatory use of face masks, so eventually some individuals wear masks, while others do not. An analogous situation was massively witnessed during the recent Olympic Games: the Brazilian volleyball player Lucas Saatkamp wore a mask during the matches, while their teammates and opponents did not. It is not a trivial matter to evaluate how well he and his teammates and opponents were protected in such scenario.

By means of controlled experimentation using affordable equipment, we evaluate the blocking capacity, as a source control and wearer protection, of three materials commonly used in home-made face masks, which are very popular in countries like Cuba. We show quantitatively that a blocking material deployed very near a source of ballistic droplets protects from them a receptor located farther away better than the same material deployed very near the receptor.

## EXPERIMENTAL

We horizontally spray blue-colored water on a screen: a nozzle producing the spray plays the role of the mouth or nose of a sneezing, coughing, or talking individual (source), while the screen plays the role of the face of a second individual (receptor). The distance between the two is $d = 17.5$ cm, which is within the inter-personal distance range observed in controlled experiments with human subjects [Zhang2020]. Between the two, an obstacle is deployed, consisting in a flat piece of porous fabric. The blocking element plays the role of a facial mask. Fig. 1 shows the main configurations used in the experiments. Since the droplets travel inside a quasi-cylindrical case, external air currents do not perturb the experiment. The nozzle is shot using two servomotors controlled by an Arduino Uno platform manually activated by a micro-switch, in order to achieve reproducibility and decrease spurious vibrations associated to direct manipulation of the nozzle. Our model experiment matches reasonably well the temporal evolution of the particle front velocity associated to coughing humans [Nishimura2013, Simha2020] (See Appendix 1).

Fig. 1(a) illustrates Configuration 1 (CONF1) that allows studying the stain pattern on the "face" of a receptor located 17.5 cm away from the source, when the former is wearing a face mask –the choice of distance is within the face-to-face proximity range measured in [Zhang2020], which can be easily experienced during rush hours in a bus in Havana, a Metro in Paris or a subway in New York. Fig. 1(b) shows Configuration 2 (CONF2), which provides the stain pattern on the "face" of the receptor located 17.5 cm away from the source, when the latter is wearing a face mask. All in all, these two configurations allow evaluating the capacity to block talking, coughing or sneezing droplets when the receiver uses a face mask, or when the source uses it, respectively. We also collected the stain patterns without any blocking material between the source and the screen in Configuration Free (CONFFREE), corresponding to Configuration 1 or 2 without any blocker.

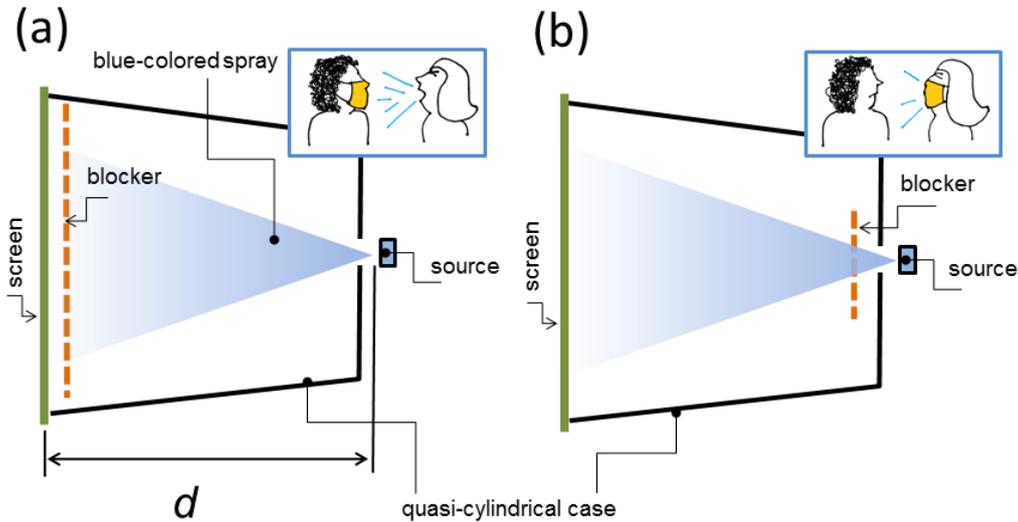

**Figure 1**. Experimental setup. (a) In Configuration 1, the screen is far from the spray source, and the blocking fabric is near the screen. (b) In Configuration 2, the screen is far from the source, but the obstacle is close to it. We call Configuration Free to either Configuration 1 or 2, without blocking material. The blue cone representing the spray is just a sketch. In (a), the blocker is 1.5 cm apart from the screen. In (b), the blocker is 4 cm apart from the source. In both (a) and (b), $d = 17.5$ cm.

Each experiment resulted in a pattern of blue stains on a white surface, which was digitally scanned and binarized, so clear and black areas correspond to places hit or not hit by droplets, respectively. Our system allows the detection of stains larger than 7 micrometers. So, we are basically detecting the stains associated to ballistic droplets, i.e., those relatively big ones expected to move as a projectile between the source and the receptor. Appendix 2 describes the experimental apparatus and procedure in more detail.

**RESULTS AND DISCUSSION**

Fig. 2 shows typical stain patterns collected around the center of the screen, after binarization. Fig. 2(a) corresponds to one CONFFREE experiment. Fig. 2(b) shows a micrograph of a table-cloth (Cotton cloth 1) used as blocking element. Fig. 2(c) and (d) correspond to experiments deploying it in the configurations CONF1 and CONF2, respectively. In our experiment, the face of the receiver is represented by the screen located 17.5 cm from the emitter (see Fig. 1), so it makes sense to compare the stain pattern shown in Fig. 2(a) (i.e., the receiver's face with no protection) with the patterns illustrated in Fig. 2(c) and Fig. 2(d), corresponding to face masks worn by the receiver and the emitter, respectively.

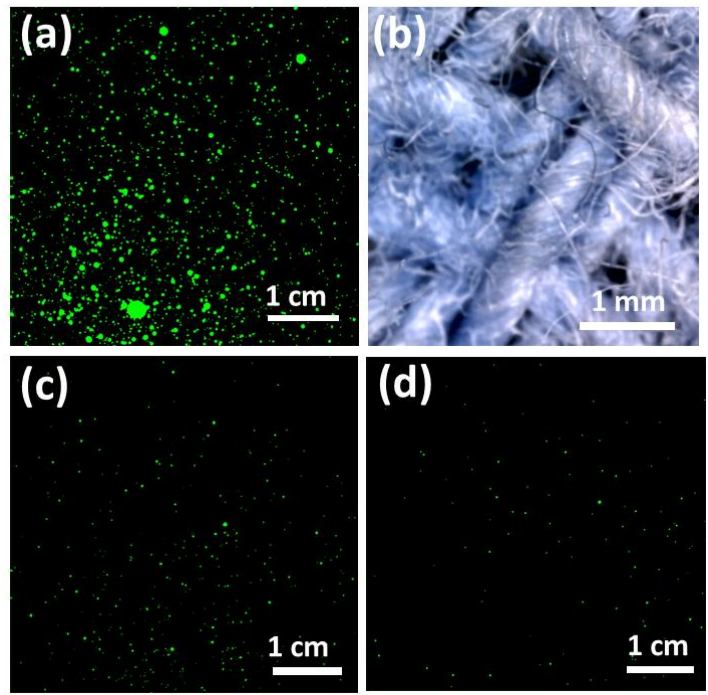

**Figure 2.** Typical stain patterns for cotton cloth 1. Image (a) shows a binarized pattern resulting from a typical experiment in CONFFREE configuration. (b) shows a micrograph of a table-cloth (Cotton cloth 1) used as blocker (See Appendix 2 for further detail). (c) and (d) are binarized patterns resulting from experiments in CONF1 and CONF2, respectively. Stains are artificially colored to aid visibility.

It is easy to see from the patterns illustrated in Fig. 2 that both CONF1 and CONF2 are able to block most of the ballistic droplets emitted by the nozzle, but a closer inspection suggests that CONF1 allows more droplets to reach the face of the receiver than CONF2.

In order to quantify the capacity of a given mask material to isolate a receptor from the ballistic contamination emitted from a source, we propose a parameter named Ballistic Blocking Capacity, given by:

$$\text{BBC} = \left( 1 - \frac{\langle \overline{\text{pixels obstacle}} \rangle}{\langle \overline{\text{pixels no obstacle}} \rangle} \right) \times 100\% \qquad (1)$$

where $\langle \overline{\text{pixels no obstacle}} \rangle$ is the average of the pixel values in the image corresponding to the screen without obstacle, averaged over the images from all similar experiments, and $\langle \overline{\text{pixels obstacle}} \rangle$ is the analogous magnitude when an obstacle is deployed. Since both averages are within the interval [0,1], BBC = 100% when the obstacle has stopped all droplets ($\langle \overline{\text{pixels obstacle}} \rangle = 0$), and BBC = 0 % when all droplets have managed to pass through the obstacle ($\langle \overline{\text{pixels obstacle}} \rangle = \langle \overline{\text{pixels no obstacle}} \rangle$).

So, for evaluating the effectiveness of a mask near the face of the receiver against ballistic droplets produced by the emitter with no mask, we insert in (1) the data from images like the ones illustrated in Fig. 2(a) and (c), averaged over all the repetitions of the experiment. For evaluating the effectiveness of the emitter's mask to protect the receiver's while the latter is not using protection, we insert in (1) the data from images associated to Fig. 2(a) and (d). The results quantitatively confirm the qualitative inspection of the patterns: formula (1), with values averaged over three repetitions of the experiments made with Cotton Cloth 1, gives BBC values of (89.82 ± 3.96) % and (97.10 ± 1.67) % when the receiver is protected by her/his own mask, and when protected by the mask used by the emitter, respectively.

**Table 1.** Ballistic Blocking Capacity for various fabrics. **Note**: See Appendix 3 for uncertainty analysis.

| Obstacle | Configuration | $\langle \overline{\text{pixels obstacle}} \rangle$ | $\langle \overline{\text{pixels no obstacle}} \rangle$ | BBC |
|---|---|---|---|---|
| Cotton cloth 1 (Table-cloth) | CONF1 | 0.004223 ± 0.000622 | | (89.82 ± 3.96) % |
| | CONF2 | 0.001203 ± 0.000540 | | (97.10 ± 1.67) % |
| Cotton cloth 2 (Handkerchief) | CONF1 | 0.002390 ± 0.002269 | | (94.24 ± 5.75) % |
| | CONF2 | 0.000157 ± 0.000130 | 0.041475 ± 0.014929 | (99.62 ± 0.34) % |
| Cotton cloth 3 (Bed sheet) | CONF1 | 0.000077 ± 0.000091 | | (99.81 ± 0.19) % |
| | CONF2 | 0.000008 ± 0.000009 | | (99.98 ± 0.02) % |
| Level 3 surgical gown (Aurora AAMI PB70) | CONF1 | 0.000006 ± 0.000005 | | (99.99 ± 0.01) % |
| | CONF2 | 0.000000 ± 0.000000 | | (100.00 ± 0.00) % |

Analogous results are found using two other cotton cloths with different average pore size and thread thickness, as well as for a Level 3 surgical gown fabric, as shown in Table 1. To test if the BBC values corresponding to CONF1 and CONF2 are statistically different for a given blocking material, a one-tailed Mann-Whitney U-test was performed. In all cases the null hypothesis was rejected within a 95% confidence level, proving that CONF2 BBC were statistically larger than those corresponding to CONF1.

**CONCLUSIONS**

By using frugal apparatus and a relatively simple experimental protocol, we have quantitatively demonstrated that porous materials commonly used in the fabrication of face masks at home are able to stop a large proportion of the liquid carried by jet droplets larger than a few microns. Such droplets are able to carry active SARS-CoV-2 virus with the potential to replicate in the nasopharynx [Wolfel2020]. Our study complements most of the current research available in the literature, which tends to concentrate on much smaller droplets (aerosols) [Konda2020, Sickbert-Bennet2020, Akhtar2020, Shah2021].

While the details of the physical mechanisms explaining the blocking capacity depend on the specific nature of the blocking material, the stain pattern protocol described here systematically reveals the differences between the near-source and near-receptor mask configurations: it is quantitatively shown that a cloth deployed very near a source of ballistic droplets protects from them a receptor located farther away better than the same cloth deployed very near the receptor. All in all, our results are consistent with the common –but hardly intuitive– belief that a face mask worn by an infected subject is effective to avoid spreading the disease, while a healthy person is less protected from an external infection source when wearing the mask, if others are not wearing it.

**Acknowledgments**

P. G. Gil is acknowledged for suggesting home-made masks as an urgent and important subject of study. T. Shinbrot contributed with key suggestions and revision of the manuscript. N. Martínez coordinated lab access a number of times during home self-isolation period. P. Altshuler-Rivera is dearly thanked for collaborating in spray visualization experiments at home.

## APPENDIX 1: Visualization of the free jet.

The jet was visualized by illuminating with a light "cone" produced by a green laser ($\lambda = 532$ nm and peak power of 50 mW). The images were acquired at 30 fps using a NIKON D5100 camera. Figure S1 illustrates the evolution of the jet based in its visualization.

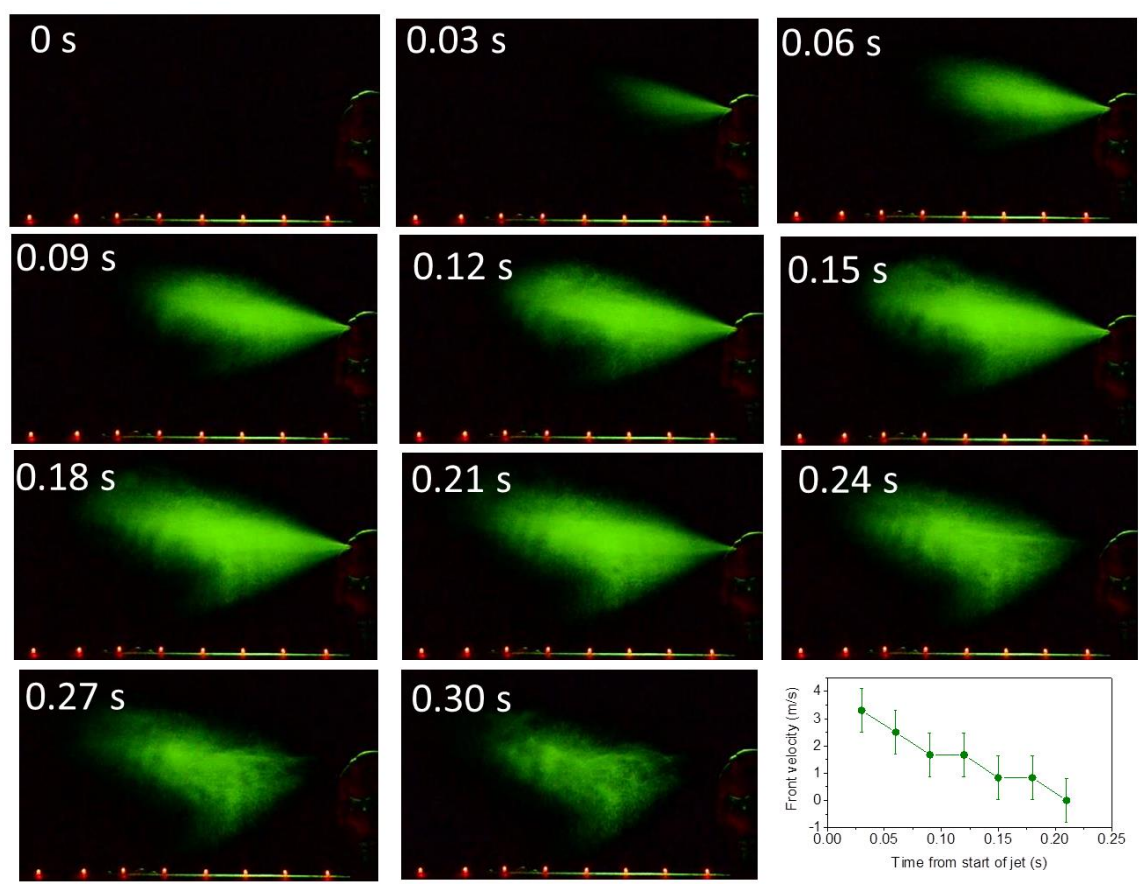

**Fig. S1 Visualization of the free jet**. The 11 photographs are a time sequence of snapshots of the free jet, visualized using laser light. The time count starts when the nozzle starts generating liquid. In each of the snapshots, the green light indicates the presence of liquid particles emitted by the nozzle (located at far right within each image). The red spots are LEDs located at a distance of 5 cm from each other. Notice that the spray can be assumed as a cone making an angle of $10^{o}$ around the horizontal direction. Bottom right panel: velocity of the jet front as time goes by.

# APPENDIX 2: EXPERIMENTAL DETAILS

### Image acquisition and processing

75 g/m$^2$-density paper sheets containing the stain patterns were scanned with a resolution of 3200 dpi and 24-bit RGB color, using a scanner EPSON model Perfection V370 Photo. Then, the image was negated, and its red component binarized. As the spots are blue, this component provided the best contrast between the stains and the background. After careful comparison between the original patterns and the binarized images, we chose a binarization threshold of 0.3.

If the image with the stain pattern was substracted from the scan of that very same sheet of paper made before the droplets were shot (in order to account for intrinsic lack of homogeneity of the paper), the results for BBC were negligibly different.

### Micrographs of blocking materials

Micrographs of different blocking materials were obtained using a Dino-lite Premier digital microscope.

# APPENDIX 3: UNCERTAINTY ANALYSIS

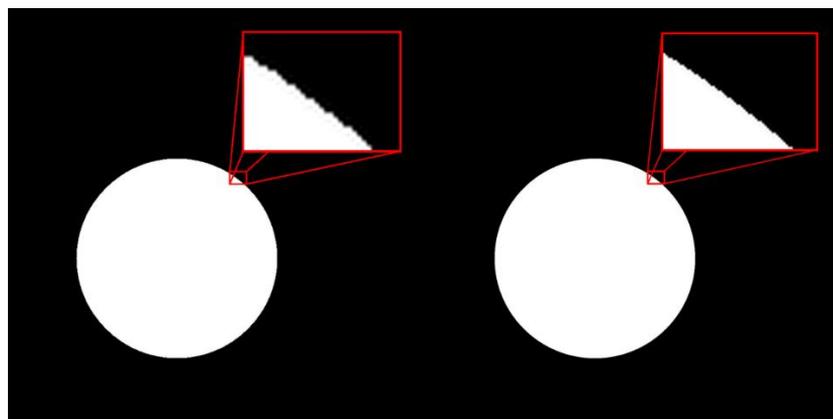

Fig. S4 Uncertainties in image digitizing. Two raster images created from the same white disk on a black background vector graphics, simulating the stains in the binarized images of the experiments. Left panel: 3200 dpi resolution image. Right panel: 6400 dpi resolution image. In the insets it can be seen the difference between the analog regions in the disks boundaries.

In the calculation of $\left\langle \overline{\text{pixels no obstacle}} \right\rangle$ and $\left\langle \overline{\text{pixels obstacle}} \right\rangle$ (both referred as *image averages* from now on, as they are computed in the same way), there are two sources of Type B standard uncertainty: the stained paper scan resolution and the finite precision with which the program reads the pixel values.

In the case of the former, a test was made to check if an increase in the image resolution would affect significantly the values of *image average* calculated in the experiments. In order to do so, two raster images were created from the same vector using resolutions of 3200 and 6400 *dpi*. The *image average* was calculated on both of them. The difference between both was around 0.1% of the value of the 3200 *dpi* image, so the influence of this source of uncertainty can be neglected as compared to the influence of the fluctuations between repetitions of the same experiment, as we will see.

The program used for image processing reads by default the pixel values with a precision of 15 significant decimal digits. As the pixel values are within the interval [0,1] the difference between two pixels whose values differ up to $R = 10^{-15}$ can be detected. Due to binarization, this resolution would only affect the pixels whose values are within the interval $[Th - R, Th + R]$ being $Th$ the binarization threshold used in the experiments. In a typical image, the proportion of pixels that satisfies this condition varies from zero to $10^{-15}$, so the influence of that source of uncertainty can be also neglected. In summary, the influence of Type B is negligible compared to the Type A standard uncertainty on the evaluation of the *image average's* combined standard uncertainty.

For all experimental configurations, *image average's* Type A standard uncertainty was computed using the well-known positive square-root of the unbiased estimator for the variance of a sample divided by its length. Each experiment was repeated 3 times and a 95% confidence interval was selected for the calculations, so according to Eq. (1), the formula used for the expanded uncertainty of BBC is

$$u(BBC) = t_2 \left( \frac{1}{\langle \overline{p_{no}} \rangle^2} u_A{}^2(\langle \overline{p_o} \rangle) + \frac{\langle \overline{p_o} \rangle^2}{\langle \overline{p_{no}} \rangle^4} u_A{}^2(\langle \overline{p_{no}} \rangle) \right)^{1/2}$$

where we have defined $\langle \overline{p_o} \rangle = \langle \overline{\text{pixels obstacle}} \rangle$, $\langle \overline{p_{no}} \rangle = \langle \overline{\text{pixels no obstacle}} \rangle$ and $t_2 = 4.30$ is the factor corresponding to the confidence interval of choice and the effective degree of freedom.